\documentclass[10pt,a4paper]{article}
\usepackage{graphicx,amsmath,amssymb,fixmath,theorem}

\newcommand{\Ord}[1]{ {\cal O}( #1 )}
\newcommand{\R}{{\Bbb R}}
\newcommand{\Z}{{\Bbb Z}}
\newcommand{\N}{{\Bbb N}}

\newcommand{\D}[2]{ \ensuremath{ \frac{d #1 }{d #2 } }}

\newcommand{\vect}[1]{\ensuremath{ \mathbold #1 } }

\theoremstyle{break}\newtheorem{theorem}{Theorem}
\theoremstyle{break}\newtheorem{definition}{Definition}

\title{Tropical geometries and dynamics of biochemical networks. Application to hybrid cell cycle models.}

 \author{V.~Noel$^1$, D.~Grigoriev$^2$, S.~Vakulenko$^3$ and O.~Radulescu$^4$ \\ \\ \small $^1$IRMAR UMR 6625, University of Rennes 1, Rennes, France, \\
  \small $^2$ CNRS, Math\'ematiques, Universit\'e de Lille, 59655,
  Villeneuve d'Ascq, France, \\
 \small  $^3$ Saint Petersburg State University of Technology and Design, St.Petersburg, Russia, \\
 \small  $^4$ DIMNP UMR CNRS 5235, University of Montpellier 2, Montpellier, France.}
\begin{document}

\maketitle

\centerline{\bf Abstract}

We use the Litvinov-Maslov correspondence principle  to reduce and
hybridize networks of biochemical reactions. We apply this method to a cell cycle oscillator model.
The reduced and hybridized model can be used as a hybrid model for the cell cycle.
We also propose a practical recipe for detecting quasi-equilibrium QE reactions
and quasi-steady state QSS species in biochemical models with rational rate functions
and use this recipe for model reduction.
Interestingly, the QE/QSS invariant manifold of the smooth model and the reduced dynamics
along this manifold can be put into correspondence to the tropical variety of the
hybridization and to sliding modes along this variety, respectively


\section{Introduction.}\label{intro}

Systems biology develops biochemical dynamic models of various cellular processes
such as signalling, metabolism, gene regulation. These models can reproduce complex
spatial and temporal dynamic behavior observed in molecular biology experiments.
In spite of their complex behavior, currently
available dynamical
models are relatively small size abstractions, containing only
tens of  variables. This modest size results from the lack of
precise information on kinetic parameters of the biochemical reactions on one hand,
and of the limitations of parameter identification methods on the other hand.
Further limitations can result from the combinatorial explosion
of interactions among molecules with multiple modifications and
interaction sites \cite{danos2007rule}.
In middle out modeling strategies small models can be justified by saying that
one looks for an optimal level of complexity that captures the salient features
of the phenomenon under study. The ability to choose the relevant details and
to omit the less important ones is part of the art of the modeler.
Beyond modeler's art, the success of simple models relies on an important
property of large dynamical systems.
The dynamics of multiscale, dissipative, large biochemical models,
can be reduced to that of simpler models, that were called dominant
subsystems \cite{radulescu2008robust,gorban2009asymptotology,gorban-dynamic}.
Simplified, dominant subsystems contain less parameters and are more easy to analyze.
The choice of the dominant subsystem depends on the comparison among the time scales
of the large model. Among the conditions leading to dominance and allowing to generate
reduced models, the most important are quasi-equilibrium (QE) and the quasi-steady state
(QSS) approximations \cite{gorban2009asymptotology}.
In nonlinear systems, timescales and together with them dominant subsystems
can change during the dynamics and undergo more or less sharp transitions.
The existence of these transitions suggests that a hybrid,
discrete/continous framework is well adapted for the description of the
dynamics of large nonlinear systems with multiple time scales
\cite{crudu2009hybrid,noel2010,noel2011}.

The notion of dominance can be exploited to obtain simpler models from larger
models with multiple separated timescales and to assemble these simpler models
into hybrid models.
This notion is asymptotic
and a natural mathematical framework to capture multiple asymptotic relations is
the tropical
geometry.
Motivated by applications
in mathematical physics \cite{litvinov1996idempotent},
systems of polynomial equations \cite{sturmfels2002solving},
etc.,  tropical geometry uses a change of
scale to transform nonlinear systems into discontinuous piecewise
linear systems. The tropicalization is a robust property of the system,
remaining constant for large domains of parameter values; it can reveal
qualitative stable features of the system's dynamics, such as
various types of attractors.
Thus, the use of
tropicalization to model large systems in molecular biology could be
a promising solution to the problem of incomplete or imprecise
information on the kinetic
parameters.

In this paper we propose a method for reduction and hybridization
of biochemical networks. This method, based on tropical geometry,
could be used to automatically produce the simple models that
are needed in middle-out approaches of systems biology.

\section{Biochemical networks with rational rate functions.}

Systems biology models use the formalism of chemical kinetics to model
dynamics of cellular processes. We consider here that the molecules of
various species are present in sufficient large numbers and that
stochastic fluctuations are negligible as a consequence of the
law of large numbers and/or of the averaging theorem \cite{crudu2009hybrid}. We also
consider that space transport phenomena are sufficiently rapid
such that the well stirred reactor hypothesis is valid. In these conditions,
the dynamics of the biochemical system can be described by systems
of differential equations. In chemical kinetics, enzymatic reactions are
often presented as indivisible entities characterized by stoichiometry vectors
and rate functions. However, each enzymatic reaction can be decomposed into
several steps that define the
reaction mechanism. The resulting stoichiometry
and global rate depend on the mechanism. Several methods were designed
for calculating effective rates of arbitrarily complex mechanisms.
For linear mechanisms  King and Altman
\cite{king1956schematic} proposed a graphical method to compute
global rates; these are rational
functions of the concentrations (an example is the Michaelis-Menten
equation). Yablovsky and Lazman \cite{lazman2008overall} studied the same problem for non-linear
mechanisms and found that in this case the reaction rates are solutions of polynomial equations; these can be solved by radicals in a few number of cases and can be
calculated by multi-variate hypergeometric series
in general \cite{lazman2008overall}. Truncation of these series to finite
order leads to rational approximations of the reaction rates.

In chemical kinetics with rational reaction rates
the concentration $x_i$ of the $i$-th component follows
the ordinary differential equation:
\begin{equation}
\D{x_i}{t} = P_i (\vect{x})/Q_i(\vect{x}),
\label{rationalsystem}
\end{equation}
where
$P_i(\vect{x}) = \sum_{\alpha \in A_i} a_{i,\alpha} \vect{x}^\alpha$,
$Q_i(\vect{x}) = \sum_{\beta \in B_i} b_{i,\beta} \vect{x}^\beta$,
are polynomials and we have $1 \leq i \leq n$. Here
$\vect{x}^\alpha = x_1^{\alpha_1} x_2^{\alpha_2} \ldots x_n^{\alpha_n}$,
$\vect{x}^\beta = x_1^{\beta_1} x_2^{\beta_2} \ldots x_n^{\beta_n}$, $a_{i,\alpha}, b_{i,\beta}$, are
nonzero real numbers, and $A_i, B_i$ are finite subsets of $\N^n$ called supports of $P_i$ and $Q_i$.

A simple example of model with rational reaction rates is the minimal
cell cycle oscillator model proposed by Tyson \cite{tyson1991modeling}.
This example
will be studied throughout the paper.
The dynamics of this nonlinear model that contains 5 species and 7 reactions
is described by a system of 5 polynomial differential equations:
\begin{eqnarray}
 y_1' & =k_9 y_2 - k_8 y_1 + k_6 y_3, \notag \\
  y_2' &=k_8 y_1 - k_9 y_2 - k_3 y_2 y_5, \notag \\
  y_3' &=k_4' y_4 + k_4 y_4 y_3^2/C^2 - k_6 y_3, \notag \\
  y_4' &= - k_4' y_4 - k_4 y_4 y_3^2/C^2 + k_3 y_2 y_5, \notag \\
  y_5' &= k_1 - k_3 y_2 y_5, \label{tyson6} \\
\text{where} & y_1 + y_2 + y_3+y_4 = C.  \notag
\end{eqnarray}


\section{Hybridization and tropical geometry.}

Tropical geometry is a new branch of  algebraic geometry that studies the asymptotic
properties of varieties. While algebraic geometry deals with polynomial functions,
tropical geometry  deals with piecewise linear functions with integer directing
slopes. Tropical geometry has a growing number of applications in enumerative
problems in nonlinear equation solving \cite{rojas2003polyhedra},
statistics \cite{pachter2004tropical}, traffic optimization
\cite{aubin2010macroscopic}.

The logarithmic transformation $u_i = log x_i,\, 1 \leq i \leq n$, well
known for drawing graphs on logarithmic paper, plays a central role in tropical geometry \cite{viro2008sixteenth}. By {\em abus de langage}, here we call {\em logarithmic paper}
the image of $\R^n_+$ by the logarithmic transformation, even if $n > 2$.
Monomials $M( \vect{x}) = a_{\alpha} \vect{x}^\alpha$ with positive coefficients $a_{\alpha}>0$,
become linear functions, $log M = log a_{\alpha} + <\alpha,log(\vect{x})>$, by this transformation.
Furthermore, the euclidian distance on the logarithmic paper is a good measure of separation (see
next section).

Litvinov and Maslov \cite{litvinov2001idempotent,litvinov1996idempotent}
proposed a heuristic (correspondence
principle) allowing to transform mathematical objects (integrals, polynomials) into
their quantified (tropical) versions. According to this heuristic, to a polynomial
with positive real coefficients  $\sum_{\alpha \in A} a_{\alpha} \vect{x}^\alpha$,
one associates the max-plus polynomial
$max_{\alpha \in A} \{log( a_{\alpha}) + < log(\vect{x}), \alpha > \}$.

We adapt this heuristic to associate a piecewise-smooth hybrid model to the systems
of rational ODEs \eqref{rationalsystem}.

\begin{definition}
We call tropicalization  of the smooth ODE
system \eqref{rationalsystem} the following piecewise-smooth system:

\begin{equation}
\D{x_i}{t} = s_i exp [max_{\alpha \in A_i} \{ log( |a_{i,\alpha}|) +
< \vect{u} , \alpha > \} - max_{\beta \in B_i} \{ log(
|b_{i,\beta}|) + < \vect{u} , \beta > \}], \label{fraction}
\end{equation}

\noindent where $\vect{u} = (log x_1,\ldots,log x_n)$, $s_i =
sign(a_{i,\alpha_{max}}) sign(b_{i,\beta_{max}})$ and
$a_{i,\alpha_{max}},\, \alpha_{max}\in A_i$ (respectively,
$b_{i,\beta_{max}},\, \beta_{max}\in B_i$) denotes the coefficient
of a monomial of the numerator (respectively, of the denominator)
for which the maximum occurring in \eqref{fraction} is attained.
\end{definition}

In a different notation this reads:

\begin{equation}
\D{x_i}{t} = Dom \{a_{i,\alpha} \vect{x}^\alpha \}_{\alpha \in A_i} / Dom \{ b_{i,\beta} \vect{x}^\beta\}_{\alpha \in B_i},
\end{equation}

\noindent
where $Dom \{a_{i,\alpha}  \vect{x}^\alpha \}_{\alpha \in A_i} =
sign(a_{i,\alpha_{max}}) exp [max_{\alpha \in A_i} \{ log( |a_{i,\alpha}|) + < \vect{u} , \alpha > \}]$.

Finally, the tropicalization can be written with Heaviside functions:

\begin{equation}
\D{x_i}{t} = \frac{\sum_{\alpha \in A_i } a_{i,\alpha} \vect{x}^\alpha \prod_{\alpha ' \ne \alpha} \theta( <\alpha-\alpha ',log(\vect{x})> +
  log(|a_{i,\alpha}|) - log(|a_{i,\alpha'}|) ) }
  {\sum_{\beta \in B_i } b_{i,\beta} \vect{x}^\beta \prod_{\beta ' \ne \beta} \theta( <\beta-\beta ',log(\vect{x})> +
  log(|b_{i,\beta}|) - log(|b_{i,\beta'}|) ) },
\end{equation}
\noindent
where $\theta(x) = 1$ if $x>0$, $0$ if not.

The following definitions are
standard and will be used throughout the paper:

\begin{definition}
The Newton polytope of a polynomial  $P(\vect{x}) = \sum_{\alpha \in A} a_{\alpha} \vect{x}^\alpha$
is defined as the convex hull of the support of $P$, $New (P) = conv(A)$.
\end{definition}

\begin{definition}
The max-plus polynomial $P^\tau(\vect{x}) = max \{ log |a_\alpha| + <\alpha,log(\vect{x})> \}$
is called the tropicalization of $P(\vect{x})$. The logarithmic function is defined as
$log(\vect{x}) : \R_+^n \to \R^n$, $log(\vect{x})_i = log(x_i)$.
\end{definition}

\begin{definition}
The set of points $\vect{x}\in\R^n$ where $P^\tau(\vect{x})$ is not smooth is called tropical
variety. Alternative names are used such as logarithmic limit sets, Bergman fans, Bieri-Groves sets,
or non Archimedean amoebas \cite{passare2005amoebas}.
\end{definition}

In two dimensions, a tropical variety is a tropical curve made of several
half-lines (tentacles) and finite intervals \cite{mikhalkin18enumerative}. A tropical line corresponds to
only three monomials and is made of three half lines sharing a common point.
The tentacles and the intervals of the tropical variety are
orthogonal to the edges and point to the interior of the Newton
polygon \cite{passare2005amoebas} (see Fig.\ref{fig1}).


\section{Dominance and separation.}

The above heuristic is related to the notion of dominance. Actually we have replaced each polynomial
in the rational function by the dominant monomial. Dominance of monomials has an asymptotic
meaning inside cones of the logarithmic paper.
For instance ${\vect{x}}^\alpha$ dominates ${\vect{x}}^\beta$ on the
 half plane $< log (\vect{x}), \alpha - \beta > > 0$ of the logarithmic paper. We have
 ${\vect{x}}^\beta/{\vect{x}}^\alpha \to 0$ when the limit is taken along lines in this half plane.

For practical applications, we would also need a finite scale notion of dominance.

Let $M_1 (\vect{x})= a_{\alpha_1} \vect{x}^{\alpha_1}$ and $M_2 (\vect{x})= a_{\alpha_2} \vect{x}^{\alpha_2}$
be two monomials. We define the following binary relations:

\begin{definition}[Separation]
$M_1$ and $M_2$ are separated on a domain $D \subset R^n_+$ at a level $\rho >0$ if
$|log (|a_{\alpha_1}| \vect{x}^{\alpha_1}) - log (|a_{\alpha_2}| \vect{x}^{\alpha_2}) | > \rho$
for all $\vect{x} \in D$.
\end{definition}

On logarithmic paper, two monomials  are separated on the domain $D$, if $D$
is separated by the euclidian distance $\rho$ from the hyperplane
$<log (\vect{x}), \alpha_1 - \alpha_2> = log |a_{\alpha_2}| -  log |a_{\alpha_1}|$.

\begin{definition}[Dominance]
The monomial $M_1$ dominates the monomial $M_2$ at the level $\rho >0$,
$M_1 \succ_\rho M_2$, if $log (|a_{\alpha_1}| \vect{x}^{\alpha_1}) > log (|a_{\alpha_2}| \vect{x}^{\alpha_2}) + \rho$ for all  $\vect{x} \in D
\subset \R_+^n$.
\end{definition}

Dominance is a partial order relation on the set of multivariate monomials defined
on subsets of $\R_+^n$.

\section{Dominance and global reduction of large models.}

There are two simple methods for model reduction of nonlinear models with multiple
timescales: the quasi-equilibrium (QE) and the quasi-steady state (QSS) approximations.
As discussed in \cite{gorban2009asymptotology}, these  two approximations are physically and
dynamically distinct. Here we present a method allowing to detect QE reactions and QSS species.

Like in \cite{radulescu2008robust}, the first step of the method is to detect
the "slaved" species, i.e. the species that obey quasi-steady state equations.
These can be formally defined by introducing the notion of imposed trace.
Given the traces $\vect{x}(t)$ of all the species, the imposed trace of the
$i$-th species is a real solution $x_i^*(t)$ of the polynomial equation $P_i(x_1(t),\ldots,x_{i-1}(t),x_i^*(t),x_{i+1}(t),\ldots,x_n(t))=0$.
Eventually, there may be several imposed traced, because a polynomial equation
can have several real solutions.

\begin{definition}
We say that
a species is slaved if the distance between
the traces $x_i(t)$ and some imposed trace $x_i^*(t)$ is small on some interval, $sup_{t \in I} |log(x_i(t))-log(x_i^*(t))| < \delta$, for some $\delta>0$ sufficiently small.
A species is globally slaved if $I = (T,\infty)$ for some $T\geq 0$.
\end{definition}

 Slaved species are good candidates for QSS species
and this criterion was used to identify QSS species in \cite{radulescu2008robust}.
More generally, slaved species are involved in rapid processes, but are not always
QSS. Actually, two distinct cases lead to slaved species.

{\em Quasi-equilibrium.}
A system with fast, quasi-equilibrium reactions has the following structure \cite{gorban2009asymptotology}:
\begin{equation}
\D{\vect{x}}{t} = \sum_{s, slow} R_s(\vect{x}) \vect{\gamma}^s + \frac{1}{\epsilon}
 \sum_{f, fast} R_f(\vect{x}) \vect{\gamma}^f, \label{QEdyn}
\end{equation}
where $\epsilon>0$ is a small parameter $\vect{\gamma}^s,\vect{\gamma}^f \in \Z^n$ are
stoichiometric vectors. The reaction rates  $R_s(\vect{x})$, $R_f(\vect{x})$
are considered rational functions of $\vect{x}$.

To separate slow/fast variables, we have to study the spaces of linear
conservation law of the initial system \eqref{QEdyn} and
of the following fast subsystem:
\begin{equation}
\D{\vect{x}}{t} = \frac{1}{\epsilon}
 \sum_{f, fast} R_f(\vect{x}) \vect{\gamma}^f. \label{fastQE}
\end{equation}

In general, the system (\ref{QEdyn}) can  have several conservation laws.
These are linear functions $b^1(\vect{x}),\ldots ,b^m(\vect{x})$ of
the concentrations that are constant in time. The conservation
laws of the system (\ref{fastQE}) provide variables that are constant on the
fast timescale. If they are also conserved by the full dynamics,
the system has no slow variables (variables are either fast or constant).
In this case, the dynamics of the fast variables is simply given
by  Eq.(\ref{fastQE}). Suppose now that the system (\ref{fastQE}) has
some more conservation laws $b^{m+1}(\vect{x}),\ldots ,b^{m+l}(\vect{x}),$
that are not conserved by the full system (\ref{QEdyn}). Then, these
provide the slow variables of the system.
The  fast variables are those $x_i$ such that $(\vect{\gamma}^f)_i  \ne 0$, for
some fast reaction $f$.

Let us suppose that the fast system \eqref{fastQE} has a stable steady state  that is
a solution of the QE equations (augmented by the conservation laws of the fast system):
\begin{eqnarray}
& \sum_{f, fast} R_f( \vect{x} ) \vect{\gamma}^f = 0, \label{QEsystem} \\
& b^i(\vect{x}) = C_i, \quad 1 \le i \le m+l.
\end{eqnarray}

By classical singular perturbation methods \cite{Tikhonov,Wasow} one can show
that the fast variables can be decomposed as $x_i = \tilde x_i + \eta_i$
where $\tilde x_i$ satisfy the QE equations \eqref{QEsystem}
 and $\eta_i = \Ord{\epsilon}$, meaning
that the fast variables $x_i$ are slaved
\cite{gorban2009asymptotology}.

Let $P_i$, $\tilde P_i$ be the numerators of the rational functions
$\sum_{s, slow} R_s(\vect{x}) \vect{\gamma}^s_i + \frac{1}{\epsilon}
 \sum_{f, fast} R_f(\vect{x})  \vect{\gamma}^f_i$ and
 $\sum_{f, fast} R_f(\vect{x})  \vect{\gamma}^f_i$, respectively. We call
 $\tilde P_i$ the pruned version of $P_i$.
 When $\epsilon$ is small enough, the monomials of the pruned version $\tilde P_i$
 dominate the monomials of $P_i$. This suggests a practical recipe for
 identifying QE reactions:

\noindent{\bf Algorithm 1}
\begin{description}
\item
Step 1: Detect slaved species.
\item
Step 2: For each $P_i$ corresponding to slaved species, compute the pruned version
 $\tilde P_i$ by eliminating all monomials
 that are dominated by other monomials of $P_i$.
\item
 Step 3: Identify, in the structure of $\tilde P_i$ the forward
 and reverse rates of QE reactions. This step could be performed by recipes presented in \cite{soliman2010unique}.
\end{description}

{\em Quasi-steady state.}
In the most usual version of QSS approximation \cite{segel1989quasi},
the species are split in two groups with concentration vectors $\vect{x}^s$
(``slow'' or basic components)
and $\vect{x}^f$ (``fast'' or QSS species).

Quasi-steady species (also called radicals or fast intermediates)
are low-concentration, slaved species. Typically, QSS species
are consumed (rather than produced) by fast reactions.
The small parameter $\epsilon$ used in singular perturbation theory
is now the ratio of small concentrations of fast intermediates to the concentration
of other species. After rescaling $\vect{x}^s$ and $\vect{x}^f$ to order one,
the set of kinetic equations reads:
\begin{eqnarray} \label{system}
\D{\vect{x}^s}{t} & = \vect{W}^s(\vect{x}^s,\vect{x}^f), \label{QSS1} \\
\D{\vect{x}^f}{t} & = (1/\epsilon) \vect{W}^f(\vect{x}^s,\vect{x}^f), \label{QSS2}
\end{eqnarray}
where the functions $\vect{W}^s$, $\vect{W}^f$ and their derivatives
are of order one ($0< \epsilon << 1$).

Let us suppose that the fast dynamics \eqref{QSS2} has a stable steady state.
The standard singular perturbation theory\cite{Tikhonov,Wasow} provides the QSS algebraic
condition $\vect{W}^f(\vect{x}^s,\vect{x}^f)=0$ which means that fast
species $\vect{x}^f$ are slaved.
These equations, together with additional balances for $x^f$
(conservation laws) are enough to deduce the fast variables  $x^f$
as functions of the slow variables $x^s$ and to eliminate
them \cite{yablonskii1991kinetic,Lazman200847,radulescu2008robust}. The slow dynamics is
given by Eq.(\ref{QSS1}).

In networks with rational reaction rates the components of
$\vect{W}^f(\vect{x}^s,\vect{x}^f)$ are rational functions. Like for QE we can define
$P_i$ as numerators of $\vect{W}^f_i$. The difference between QSS conditions with respect to QE situation
is that in the pruned polynomial $\tilde P_i$  one can no longer find forward and backward rates
of QE reactions, ie the step 3 of Algorithm 1 will
not identify reversible reactions. Alternatively, one can realize that slaved species
can have relatively large concentrations, in which case they are not QSS species.
However, it is difficult to say which concentration value separates QSS from non QSS
species among slaved species, hence the former, qualitative criterion is better.

\section{Sliding modes of the tropicalization.}

A notable phenomenon resulting from tropicalization is the occurrence of
sliding modes. Sliding modes are well known for ordinary differential
equations with discontinuous vector fields \cite{filippov1988differential}.
In such systems, the dynamics can follow discontinuity hypersurfaces
where the vector field is not defined.

The conditions for the existence of sliding modes are generally intricate. However,
when the discontinuity hypersurfaces are smooth and $n-1$ dimensional ($n$ is the
dimension of the vector field) then the conditions for sliding modes read:
\begin{equation}
<n_+(x), f_+(x)> < 0, \quad <n_-(x), f_-(x)> < 0, \quad x \in \Sigma,
\label{slidingmode}
\end{equation}
where $f_+,f_-$ are the vector fields on the two sides of $\Sigma$ and
$n_+= -n_-$ are the interior normals.

Let us consider that the smooth system \eqref{rationalsystem} has quasi-steady state species or quasi-equilibrium reactions. In this case, the fast dynamics reads:
\begin{equation}
\D{x_i}{t} = \frac{1}{\epsilon} \tilde P_i (\vect{x})/ \tilde Q_i(\vect{x}),\quad i \quad \text{fast},
\label{fastdynamics}
\end{equation}
where $\tilde P_i (\vect{x})$, $\tilde Q_i(\vect{x})$ are pruned versions of $P_i$, $Q_i$,
and $\epsilon$ is the small, singular perturbation parameter.

For sufficiently large times, the fast variables satisfy (to $\Ord{\epsilon}$):
\begin{equation}
\tilde P_i (\vect{x}) =0,\quad  i \quad \text{fast}.
\label{fasteq}
\end{equation}

The pruned polynomial is usually a fewnomial (contains a small number
of monomials). In particular, let us consider the case when only
two monomials remain after pruning,
$\tilde P_i(\vect{x}) =  a_{1} \vect{x}^{\alpha_1} + a_{2} \vect{x}^{\alpha_2}$.
Then, the equation \eqref{fasteq} defines
a hyperplane $S = \{ <log(\vect{x}),\alpha_1-\alpha_2> = log (|a_{1}|/|a_{2}|) \}$.
This hyperplane belongs to the tropical variety of $\tilde P_i$,
because it is the place where the monomial $\vect{x}^{\alpha_1}$ switches
to $\vect{x}^{\alpha_2}$ in  the max-plus polynomial defined by $\tilde P_i$.
For $\epsilon$ small, the QE of QSS conditions guarantee the existence of an
invariant manifold ${\mathcal M}_\epsilon$, whose distance to $S$ is $\Ord{\epsilon}$.

Let $n_+,n_-$ defined as above
and let
$(f_+)_i = \frac{1}{ \tilde Q_i(\vect{x})} a_{1} \vect{x}^{\alpha_1}  $,
$(f_-)_i = \frac{1}{ \tilde Q_i(\vect{x})} a_{2} \vect{x}^{\alpha_2}  $,
$f_i = \frac{1}{\epsilon} [(f_+)_i+(f_-)_i]$
for $i$ fast, $(f_+)_j=(f_-)_j=f_j=\frac{\tilde P_j}{ \tilde Q_j}$, for $j$ not fast.
Then, the stability conditions for the invariant manifold
read $ <n_+(x_+), f(x_+)> < 0$, $ <n_-(x_-), f(x_-))> < 0$, where
$x_+,x_-$ are close to ${\mathcal M}_\epsilon$ on the side towards
which points $n_+$ and $n_-$, respectively.
We note that
$|(f_+)_i (x_+)| > |(f_-)_i (x_+)|$.
Thus,
$ <n_+, f> = \frac{1}{\epsilon} (n_+)_i [(f_+)_i  +  (f_-)_i]
  + \sum_{j, not fast} (n_+)_j (f_+)_j$
and
$ <n_+, f_+>  = \frac{1}{\epsilon} (n_+)_i (f_+)_i  + \sum_{j, not fast} (n_+)_j (f_+)_j$.
Thus, if $<n_+, f> < 0$, then for $\epsilon$ small enough
$(n_+)_i (f_+)_i <0$ and $<n_+, f_+> < 0$ because  $<n_+, f> > <n_+, f_+>$.
Similarly, we show that $<n_-, f> < 0$ implies $<n_-, f_-> < 0$.
This proves the following
\begin{theorem}
If the smooth dynamics obeys QE or QSS conditions and if the pruned polynomial
$\tilde P$ defining the fast dynamics is a 2-nomial, then the QE or QSS equations
define a hyperplane of the tropical variety of $\tilde P$. The stability
of the QE of QSS manifold implies the existence of a sliding mode of the
tropicalization along this hyperplane.
\end{theorem}

The converse result, i.e.
deducing the stability of the QE/QSS manifold from
the existence of a sliding mode on the tropical variety may be wrong.
Indeed, it is possible for a trajectory
of the smooth system to be close to a
hyperplane of the tropical variety carrying a
sliding mode and where the QE/QSS equations are satisfied identically.
However, as we will see in the next section, this
trajectory can leave the hyperplane
sooner than the sliding mode.











\begin{figure}[h!]
\begin{centering}
\includegraphics[width=80mm]{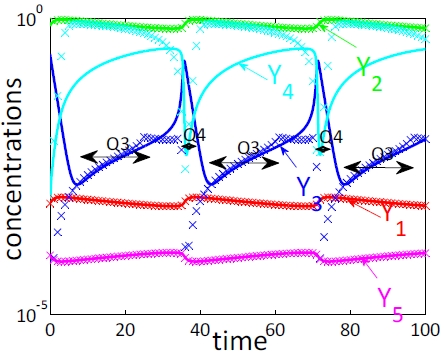}\includegraphics[width=60mm]{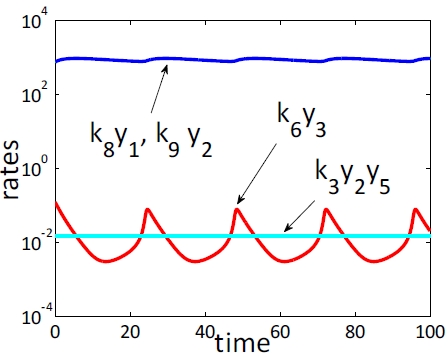} \\
\includegraphics[width=60mm]{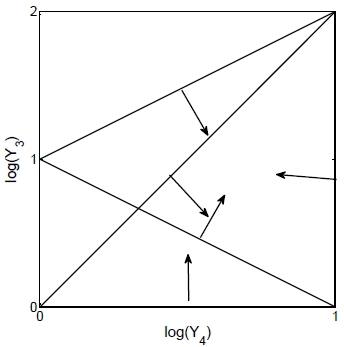} \includegraphics[width=80mm]{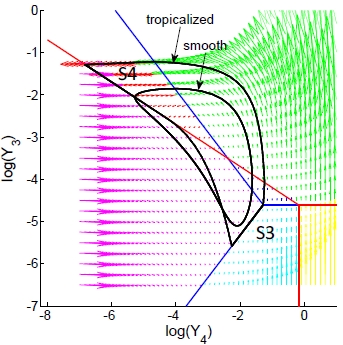}
\end{centering}
\caption{(top left) Detection of slaved species by comparing traces to
imposed traces: the species $y_1,y_2,y_5$ are slaved globally, the species
$y_3,y_4$ are slaved on intervals $Q_3$,$Q_4$, respectively. (top right) Comparison of
monomials of the polynomial systems of quasi-steady state equations.
(bottom left) Newton polygons and inner normals of the reduced two dimensional
polynomial model. (bottom right) Phase portrait on logarithmic paper
of the reduced two dimensional model. We represent the two tropical curves (the tripods graphs, a red and a blue one), the modes (smooth vector fields within domains bordered
by tropical curves tentacles), the
smooth and tropicalized limit cycles.
The tropicalized cycle contains two sliding modes $S_3$,$S_4$ corresponding
to the intervals $Q_3$, $Q_4$ on which $y_3$, $y_4$ are quasi-stationary, respectively.
\label{fig1}}
\end{figure}

\section{From smooth to hybrid models via reduction.}

Starting with the system  \eqref{tyson6} we first reduce it to
a simpler model. The analysis of the model is performed for the
values of parameters from \cite{tyson1991modeling}, namely
$k_1=0.015,k_3=200,k_4=180,k_4'=0.018,k_6=1,k_7=0.6,k_8=1000000,k_9=1000$;

In order to do that we generate one or several traces (trajectories) $y_i(t)$.
The smooth system has a stable periodic trace which
is a limit cycle attractor.
We also compute the imposed traces $y_i^*(t)$ that are solutions of the
equations:
\begin{eqnarray}
 k_9 y_2(t) - k_8 y_1^*(t) + k_6 y_3(t) = 0, \notag\\
 k_8 y_1(t) - k_9 y_2^*(t) - k_3 y_2(t) y_5(t) = 0, \notag\\
 k_4' y_4(t) + k_4 y_4(t) y_3^{*2}(t)/C^2 - k_6 y_3^*(t) = 0, \notag\\
 - k_4' y_4^*(t) - k_4 y_4^*(t) y_3^2(t)/C^2 + k_3 y_2(t) y_5(t) = 0, \notag\\
  k_1 - k_3 y_2(t) y_5^*(t) = 0.
\end{eqnarray}

We find that, for three species $y_1$,$y_2$, and $y_5$,  the distance between the traces $y_i^*(t)$ and $y_i(t)$
is small for all times which means that these species are slaved on the whole limit cycle (Figure \ref{fig1} top left).
Also, we have a global conservation law $y_1+y_2+y_3+y_4=C$, that can be obtained by summing
the first four differential equations in \eqref{tyson6}. The three quasi-steady state equations
for the three slaved species have to be solved jointly with the global conservation law:
\begin{eqnarray}
 &k_9 y_2 - k_8 y_1 + k_6 y_3 =0, \notag \\
 &k_8 y_1 - k_9 y_2 - k_3 y_2 y_5 =0, \notag \\
 &k_1 - k_3 y_2 y_5 =0, \notag \\
&y_1 + y_2 + y_3+y_4 = C.
\label{notpruned}
\end{eqnarray}

Comparison of the monomials (for values of parameters as above)
in this system shows that $ max(k_8 y_1,k_9 y_2) \succ k_6 y_3 $, and
$ max(k_8 y_1,k_9 y_2) \succ k_3 y_2 y_5$ (Fig.\ref{fig1} top right)
which leads to the pruned system:
\begin{eqnarray}
&k_8 y_1 - k_9 y_2  =0, \notag \\
 &k_8 y_1 - k_9 y_2  =0, \notag \\
 &k_1 - k_3 y_2 y_5 =0, \notag \\
&y_1 + y_2 + y_3+y_4 = C. \label{pruned}
\end{eqnarray}

The first two equations are identical and correspond to quasi-equilibrium of the reaction
between $y_1$ and $y_2$. The third equation means that $y_5$ is a quasi-steady state species.
The pruned system allows the elimination of the variables $y_1,y_2,y_5$. The slow
variable  $y_{12} = y_1 + y_2$ demanded by the quasi-equilibrium condition (this is a conservation
law of the fast system) can be eliminated
by using the global conservation law.

We note that the dominance relations leading to the pruned equations were found numerically
in a neighborhood of the periodic trace. This means that
QE and QSS approximations are valid at least on the limit cycle. More global testing
of these relations will be presented elsewhere. Note that the system \eqref{notpruned}
can be solved also without pruning. However,  \eqref{notpruned} has four
independent equations allowing to eliminate four of the five dynamic variables
leading to a one dimensional dynamical system. It turns out that the correct application
of the QE and QSS approximations has to use \eqref{pruned} and not \eqref{notpruned}.

After elimination, we obtain the following reduced differential-algebraic dynamical system:
\begin{eqnarray}
  y_3' & = k_4' y_4 + k_4 y_4 y_3^2/C^2 - k_6 y_3, \notag \\
  y_4' & = - k_4' y_4 - k_4 y_4 y_3^2/C^2 + k_1,   \notag \\
  y_1 &= (C - y_3 - y_4) k_9/(k_8+k_9), \notag  \\
  y_2 &= (C - y_3 - y_4) k_8/(k_8+k_9), \notag  \\
  y_5 &= k_1(k_8+k_9)/(k_3 k_8 (C - y_3 - y_4). \label{reduced}
\end{eqnarray}

Now we tropicalize this reduced system. The tropicalization could have been
done on the initial system in which case the pruned equations \eqref{pruned}
would indicate that the reduced dynamics is a sliding mode of the tropicalized
system on the two dimensional hypersurface $k_8 y_1 = k_9 y_2,k_1 = k_3 y_2 y_5,
y_1 + y_2 + y_3+y_4 = C$. However, although the result (concerning the dynamics
on the QE/QSS manifold) should be the same, it is much
handier to tropicalize the reduced
system \eqref{reduced}. Indeed, the tropicalization of the full 5D system is difficult to
visualize and would also produce complex modes that can not be reduced to 2D (these modes
describe the fast relaxation to the QE/QSS manifold).

The resulting hybrid model reads:
\begin{eqnarray}
  y_3' & = Dom \{ k_4' y_4 , k_4 y_4 y_3^2/C^2, - k_6 y_3\},  \notag \\
  y_4' & = Dom \{ - k_4' y_4,  - k_4  y_4 y_3^2/C^2, k_1 \}, \notag \\
\end{eqnarray}
or equivalently using Heaviside functions:
\begin{eqnarray}
  y_3' & =  k_4' y_4  \theta(-h_1 - 2 u_3 )
  \theta( h_2 + u_4 - u_3 )
  +  \frac{k_4}{C^2} y_4 y_3^2 \theta(h_1 + 2 u_3 )
  \theta(h_1+h_2 + u_4 + u_3) \notag \\
 & - k_6 y_3 \theta( -h_2 - u_4 + u_3)
  \theta(-h_1-h_2 - u_4 - u_3),
    \notag \\
  y_4' & = - k_4' y_4  \theta(-h_3 - 2 u_3 )\theta( -h_4  + u_4 )
   - \frac{k_4}{C^2}  y_4 y_3^2 \theta( h_3 + 2 u_3 )\theta( h_3 - h_4 + 2u_3 + u_4 ) \notag \\
  &  k_1  \theta(h_4  - u_4 )\theta( -h_3+h_4 - 2u_3 - u_4 ),
\end{eqnarray}
where $h_1=h_3=log( k_4/(k_4'C^2))$, $h_2=log(k_4'/k_6)$,
$h_4=log(k_1/k_4')$.

The Newton polygons of the polynomials $k_4' y_4 + k_4 y_4 y_3^2/C^2 - k_6 y_3$
and $- k_4' y_4 - k_4 y_4 y_3^2/C^2 - k_6 y_3$ are triangles
 (Fig.\ref{fig1} bottom left). The two triangles share a common edge which is
 a consequence of the fact that the reduced model have two reactions each one acting
 on the two species.
The tentacles of the two tropical curves (in red and blue in Fig.\ref{fig1} bottom right)
point in the same directions as the inner normals
to the edges of the Newton polygons (the corresponding equations  are $h_1 + 2 u_3 = 0$,
$h_2 + u_4 - u_3=0$, $h_1+h_2 + u_4 + u_3=0$ for one and
$h_3 + 2 u_3=0$,  $h_4  + u_4=0$, $h_3-h_4 + 2u_3 + u_4=0$ for the other).
These tentacles (half lines) decompose the positive quarter plane into 6 sectors corresponding to
the 6 modes of the hybrid model.

In Fig.\ref{fig1} bottom right we have also represented the phase portrait
of the reduced model on logarithmic paper. The dynamical variables are
$u_3=log(y_3)$ and $u_4=log(y_4)$. The vector field corresponding to
$u_3' = y_3'/y_3$ and $u_4' = y_4'/y_4$ was computed with the dominant
monomials in each plane sector as follows:
\begin{eqnarray}
u_4' &= - k_4 y_3^2, u_3'= - k_6 \text{ for the mode 1}, \notag \\
u_4'&= - k_4 y_3^2, u_3'=  k_4 y_3 y_4 \text{ for the mode 2}, \notag \\
u_4'&=  k_1 y_4^{-1}, u_3'=  k_4 y_3 y_4 \text{ for the mode 3}, \notag \\
u_4'&=  k_1 y_4^{-1}, u_3'= k_4' y_4 y_3^{-1} \text{ for the mode 4}, \notag \\
u_4'&=  k_1 y_4^{-1}, u_3'= - k_6 \text{ for the mode 5}, \notag \\
u_4'&= - k_4', u_3'= k_4' y_4 y_3^{-1} \text{ for the mode 6}.
\label{modes}
\end{eqnarray}

Like the smooth system, the tropicalization has a stable periodic trajectory (limit cycle).
This is represented together with the limit cycle trajectory of the smooth system
in Fig.\ref{fig1} bottom right. The period of the tropicalized limit cycle is slightly
changed with respect to the period of the smooth cycle. However, we can modulate the period
of the tropicalized cycle and make it fit the period of the smooth cycle by
acting on the moments of the mode change. This stands to displacing the tentacles of the tropical
varieties parallel to the initial positions or equivalently, to changing the parameters
$h_1,h_2,h_3,h_4$ while keeping $h_1=h_3$ which is a symmetry of the problem.

The tropicalized system has piecewise smooth hybrid dynamics. Typically, it passes
from one type of smooth dynamics (mode) described by one set of differential equations
to another smooth dynamics (mode) described by another set of differential equations
(the possible modes are listed in Eq.\eqref{modes}). The command to change the mode is intrinsic and happens when the
trajectory attains the tropical curve. However, if the sliding mode
condition \eqref{slidingmode} is fulfilled the trajectory continues along some
tropical curve
tentacle  instead of changing plane sector and evolve according to one of the
interior modes \eqref{modes}.
The tropicalized limit cycle has two sliding modes ($S_4$ and $S_3$
in Fig.\ref{fig1}). The first one is along
the half-line $h_3-h_4 + 2u_3 + u_4=0$ on the logarithmic paper
(tentacle $S_4$ on the red tropical curve in Fig.\ref{fig1}).
In order to check  \eqref{slidingmode}
we note that $f^+=(k_1 y_4^{-1},-k_6)$, $f^-=(-k_4 y_3^{2}, -k_6)$, $n^+=-n^-=(-1,-2)$.
We have a sliding mode if $- k_1 y_4^{-1} + 2 k_6 <0$, meaning that the exit from the
sliding mode occurs when $u_4 > log(k_1/(2 k_6))$.
The second sliding mode is along the tentacle
$h_2 + u_4 - u_3=0$ ($S_3$ on the blue tropical curve in Fig.\ref{fig1}).
We have $f^+=(k_1 y_4^{-1},-k_6)$, $f^-=(k_1 y_4^{-1},k_4'y_4y_3^{-1})$, $n^+=-n^-=(-1,1)$.
The conditions \eqref{slidingmode} are fulfilled when $k_1y_4^{-1}-k_4'y_4y_3^{-1}<0$ which
is satisfied on the entire tentacle. The exit
from this second mode occurs at the end of the blue tropical curve tentacle.
Interestingly, the sliding modes of the tropicalization can be put into correspondence
with places on the smooth limit cycle where the smooth limit cycle acquires new QSS species.
This can be seen in Fig.\ref{fig1} top left. The species $y_3$ becomes quasi-stationary
on time intervals $Q_3$ that satisfy (with good approximation) the relation $h_2 + u_4 - u_3=0$
and correspond to the sliding mode on the blue tropical curve. Also, the species $y_4$ becomes
quasi-stationary on very short time intervals $Q_4$ that satisfy $h_3-h_4 + 2u_3 + u_4=0$
and correspond to the sliding mode on the red tropical curve.
As pointed out in the preceding section, the trajectories of the
smooth dynamics can evolve close to the tentacles, but leave them sooner
than the sliding modes.

We end this section with a study of the bifurcations of the ODE model and of
its tropicalization. It is easy to check that there is only one degree of
freedom describing the relative position of the two tropical curves. This
is the distance between the origins of the tropical curves,
that is given by the combination $k_1 k_4'^{-1/2}k_4^{1/2} k_6^{-1}$. Thus, by changing any one of the
parameters $k_1,k_4',k_4,k_6$ we can invert the relative position of the tropical curves and change the
partition of the logarithmic paper into domains. This leads to two
Hopf bifurcations of the ODE model and also two
Hopf bifurcations of the tropicalization. The bifurcation of the tropicalization
is discontinuous and can also be delayed with respect to the continuous bifurcation
of the ODE model (Fig.2).



\begin{figure}[h!]
\begin{centering}
\includegraphics[width=60mm]{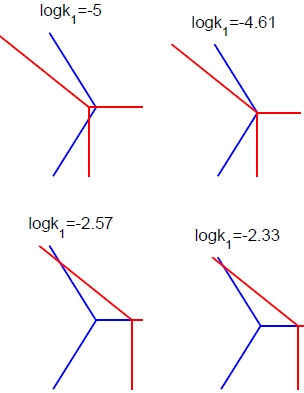}\includegraphics[width=80mm]{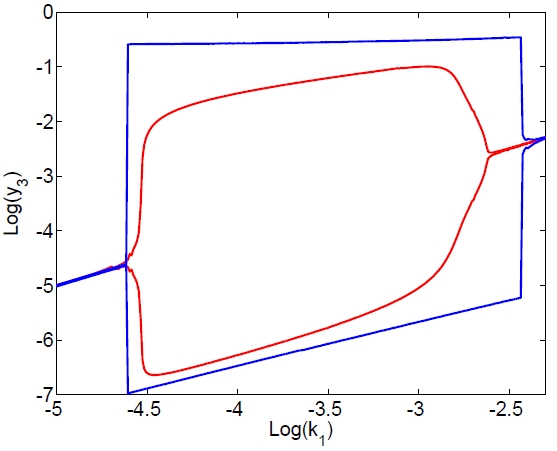} \\
\end{centering}
\caption{Hopf bifurcations of the smooth and tropicalized system. (left)
The relative positions of the tropical curves can be changed by changing the combination
$k_1 k_4'^{-1/2}k_4^{1/2} k_6^{-1}$. The first Hopf bifurcation corresponds to $k_1 k_4'^{-1/2}k_4^{1/2} k_6^{-1}=1$, i.e. $log(k_1)=-4.61$, when the tropical curves
intersect in a single point. For the second Hopf bifurcation
the relative position of the two tropical curves
is no longer exceptional; the position of the bifurcation results
from sliding modes stability analysis.
(right)
Amplitudes of oscillation are shown for the tropicalization (red)
and for the smooth system (blue);
\label{fig2}}
\end{figure}

\section{Solving ordinary differential equations in
triangular form}

We give a digest of a general algorithm for solving systems of the type
\eqref{rationalsystem} and more generally, an arbitrary system of
ordinary differential equations:
\begin{equation}
G_j(x_1,x_1^{(1)},\dots,x_1^r,x_2,x_2^{(1)},\dots,x_2^{(r)},\dots,x_n,x_n^{(1)},\dots,x_n^{(r)},t)=0,\,
1\le j\le N, \label{ordinary}
\end{equation}
where $G_j$ are differential polynomials of the {\it order} at most
$r$ in the derivatives $x_i^{(s)}=\partial ^s x_i/ \partial t^s, \,
s\le r$. Let the {\it degrees} of the differential polynomials $G_j$
do not exceed $d$. Finally, for algorithmic complexity purposes we
assume that the coefficients of $G_j$ are integers with absolute
values less than $2^l$, the latter means that the {\it bit-size of
the coefficients} $l(G_j)\le l$.

In \cite{seidenberg1956} an algorithm was designed which works not
only for ordinary differential systems like \eqref{ordinary}, but
even for systems of {\it partial} differential equations. For
ordinary systems \eqref{ordinary} the algorithm was improved in
\cite{grigoriev1989}, although still its complexity is rather big
(see below). We describe the ingredients of the output (which has a
triangular form) of the latter improved algorithm and provide for it
the complexity bounds.

The algorithm executes the consecutive elimination of the
indeterminates $x_n,\dots,x_1$. The algorithm yields a partition
$P=\{P_i\}_{1\le i\le M}$ of the space of the possible functions
$x_1$. Each $P_i$ is given by a system of an equation
$f_{i,1}(x_1,t)=0$ and an inequality $g_{i,1}(x_1,t)\neq 0$ for
suitable differential polynomials $f_{i,1},\, g_{i,1}$. Then the
algorithm yields an equation $f_{i,2}(x_1,x_2,t)=0$ and an
inequality $g_{i,2}(x_1,x_2,t)\neq 0$ for $x_2$ for suitable
differential polynomials $f_{i,2},\, g_{i,2}$. We underline that the
latter equation and inequality hold on $P_i$. One can treat the
system $f_{i,2}=0,\, g_{i,2}\neq 0$ as the conditions on $x_2$ with
the coefficients being some differential polynomials in $x_1$
(satisfying $P_i$).

Continuing in a similar way, the algorithm produces a triangular
system of differential polynomials $f_{i,3}(x_1,x_2,x_3,t)$,
$g_{i,3}(x_1,x_2,x_3,t),\dots$,$f_{i,n}(x_1,\dots,x_n,t)$,
$g_{i,n}(x_1,\dots,x_n,t)$. Thus, at the end $x_n$ satisfies (on
$P_i$) the equation $f_{i,n}(x_1,\dots,x_n,t)=0$ and the inequality
$g_{i,n}(x_1,\dots,x_n,t)\neq 0$ treated as a system with the
coefficients being differential polynomials in $x_1,\dots,x_{n-1}$.

In other words, suppose that one has a device being able to solve an
ordinary differential system $f(x)=0,\, g(x)\neq 0$ in a single
indeterminate $x$. Then the algorithm would allow one to solve the
system \eqref{ordinary} consecutively: first producing $x_1$
satisfying $f_{i,1}(x_1,t)=0,\, g_{i,1}(x_1,t)\neq 0$, after that
producing $x_2$ satisfying $f_{i,2}(x_1,x_2,t)=0,\,
g_{i,2}(x_1,x_2,t)\neq 0$ and so on.

This completes the description of the output of the algorithm. Now
we turn to the issue of its complexity. One can bound the orders of
the differential polynomials $ord(f_{i,s}),\, ord(g_{i,s})\le r\cdot
2^n\, := R,\, 1\le i\le M,\, 1\le s\le n$, the number of the
elements in the partition and the degrees $M,\, deg(f_{i,s}),\,
deg(g_{i,s})\le (Nd)^{2^R}\, :=Q$. Finally, the bit-size of the
integer coefficients of $f_{i,s},\, g_{i,s}$ and the complexity of
the algorithm can be bounded by a certain polynomial in $l,\, Q$.

Thus, the number $n$ of the indeterminates brings the main
contribution into the complexity bound, which is triple exponential in $n$.
Of course, the
above bounds have an a priori nature: they take into the account all
the conceivable possibilities in the worst case, whereas in
practical computations considerable simplifications are usually
expected.

This illustrates the gain that one can obtain from using tropical geometry
to guide model reduction and obtain systems with smaller numbers
of indeterminates.

\section{Conclusion.}

Tropical geometry offers a natural framework to study biochemical networks with
multiple timescales and rational reaction rate functions.
First, and probably most importantly,
 tropicalization can guide
model reduction of ODE systems.
We have shown that the existence of quasi-equilibrium reactions
and of quasi-stationary species implies the existence of sliding
modes along the tropical variety. Conversely, when the tropicalization has
sliding modes along hyperplanes defined by the equality of two monomials,
we propose an algorithm to decide whether the system has quasi-equilibrium
reactions or quasi-equilibrium species. This distinction allows correct
model reduction, and represents an improvement of methods
proposed in \cite{radulescu2008robust}.

The tropicalization represents an abstraction of the ODE model. This abstraction
may be not sound for some dynamic properties, but may conserve others. If the trajectories
of the ODE model are either very far or very close to the tropical
varieties, they tend to remain close
to the trajectories of the tropicalization for a while. However,
the quality of the approximation is
not guaranteed at finite distance from the tropical variety.
For instance, the exit of tropicalized trajectories
from a sliding mode tends to be delayed, and
smooth trajectories leave earlier neighborhoods of
tropical varieties.
The example studied in this paper also
illustrates some
properties of bifurcations of the tropicalization, that we have tested
numerically.
The tropicalization qualitatively preserves the
type and stability of attractors, but can also
introduce delays of a Hopf bifurcation. Thus, the
tropicalization can only roughly indicate the
position of the bifurcation of the ODE model.
Furthermore,
for Hopf bifurcations, the amplitude of the oscillations
behaves differently for the ODE model and for the
tropicalization. In fact, Hopf bifurcations are continuous
for the ODE model and discontinuous
for the tropicalization.


The tropicalization provides
in the same time a reduced model and a "skeleton" for the hybrid dynamics of the
reduced model. This skeleton, specified by the tropical varieties, is robust.
As a matter of fact,  monomials of parameters are generically
well separated \cite{gorban-dynamic}.
This implies that tropicalized and
smooth trajectories are not that far one from another.
Furthermore, because the tropicalized dynamics is robust, it
follows that the system can tolerate large relative
changes of the parameters without strong modifications of its
dynamics.

The dynamics of the model studied in this paper is relatively simple: it has a
limit cycle embedded in a two dimensional invariant manifold. As future work we intend
to extend the approach to more complex attractors, such as cycles in
dimension larger than two and chaotic attractors.
Methods to compute tropical varieties in any dimension
are well developed in tropical algebraic geometry \cite{bogart2007computing}.
Given the tropical variety, the existence of sliding modes can be easily checked and the pruned
polynomials defining the fast dynamics calculated. This should lead directly to
identification of quasi-equilibrium reactions and quasi-stationary species, without
the need of simulation (replaces Step 1 in the Algorithm 1).
Proposing simplified descriptions of the dynamics of large and imprecise systems,
tropical geometry techniques could find a wide range of applications from synthetic
biology design to understanding emerging properties of complex biochemical networks.

\section*{Aknowlegments}
VN was supported by University of Rennes 1. SV was supported by the Russian Foundation for Basic Research (Grant Nos. 10-01-00627 s and 10-01-00814 a) and the CDRF NIH (Grant No. RR07801) and by a visiting professorship grant from the University of Montpellier 2.

{\small
\newcommand{\etalchar}[1]{$^{#1}$}

\bibliographystyle{alpha}
}

\end{document}